\documentclass[12pt,preprint]{aastex}

\shorttitle{TRIS: II - CMB temperature} \shortauthors{Gervasi et
al.}

\begin{document}



\title{TRIS II: \\ search for CMB
  spectral distortions at 0.60, 0.82 and 2.5 GHz}


\author{M. Gervasi\altaffilmark{1,2}, M. Zannoni\altaffilmark{1}, A. Tartari,
G. Boella\altaffilmark{2}, and G. Sironi\altaffilmark{1}}
\affil{Physics Department, University of Milano Bicocca, Piazza
della Scienza 3, I20126 Milano Italy}

\email{mario.zannoni@mib.infn.it}


\altaffiltext{1}{also Italian National Institute for Astrophysics,
INAF, Milano.}

\altaffiltext{2}{also Italian National Institute for Nuclear
Physics, INFN, Milano-Bicocca.}


\begin{abstract}
With the TRIS experiment we have performed absolute measurements
of the sky brightness in a sky circle at $\delta = +42^{\circ}$ at
the frequencies $\nu =$ 0.60, 0.82 and 2.5 GHz. In this paper we
discuss the techniques used to separate the different
contributions to the sky emission and give an evaluation of the
absolute temperature of the Cosmic Microwave Background. For the
black-body temperature of the CMB we get: $T_{cmb}^{th}=(2.837 \pm
0.129 \pm 0.066)K$ at $\nu=0.60$ GHz; $T_{cmb}^{th}=(2.803 \pm
0.051 \ ^{+0.430} _{-0.300})K$ at $\nu=0.82$ GHz;
$T_{cmb}^{th}=(2.516 \pm 0.139 \pm 0.284)K$ at $\nu=2.5$ GHz. The
first error bar is statistic (1$\sigma$) while the second one is
systematic. These results represent a significant improvement with
respect to the previous measurements. We have also set new limits
to the free-free distortions, $ -6.3 \times 10^{-6} < Y_{ff} <
12.6 \times 10^{-6}$, and slightly improved the Bose-Einstein
upper limit, $|\mu| < 6 \times 10^{-5}$, both at 95\% confidence
level.

\end{abstract}


\keywords{cosmology: cosmic microwave background, cosmological
parameters, observations, diffuse radiation, radio continuum:
galaxies}

\section{Introduction}\label{Introduction}

At decimetric wavelenghts the sky brightness temperature
($T_{sky}$) can be written as the sum of different contributions:
cosmic microwave background ($T_{cmb}$), galactic emission
($T_{gal}$), unresolved extragalactic radio sources ($T_{uers}$).

The temperature of the cosmic microwave background (CMB),
$T_{cmb}$, can be considered, for our purposes, as isotropic. The
largest anisotropy component is the dipole term $\Delta
T_{dipole}(\alpha,\delta) = T_d \ \cos \theta$, with $ T_d = 3.381
\pm 0.007$ mK, and $\theta$ angle between the direction of
observation and the maximum of the dipole at ($\alpha = 11^h
12^m.2 \pm 0^m.8$; $\delta = -7^{\circ}.06 \pm 0^{\circ}.16$)
(\cite[]{Bennet_96}; \cite[]{Fixen_96} and \cite[]{Fixen_02}). It
gives a level of anisotropy well inside the error budget of our
absolute measurements. Therefore we will consider in the present
paper only the possible dependence of the CMB brightness
temperature on the frequency ($T_{cmb}(\nu)$). The frequency
dependence could be related to the spectral distortion of the CMB.

The most accurate measurement of the CMB  temperature so far made
is the one obtained by the COBE-FIRAS team (\cite[]{Mather_90};
\cite[]{Mather_94}; \cite[]{Fixen_96}; \cite[]{Mather_99};
\cite[]{Fixen_02}). Between 60 to 600 GHz they found a spectrum
fully compatible with a black body emitting at the thermodynamic
temperature $T_{cmb}^{th}=2.725 \pm 0.001 \ K$ (see
\cite[]{Fixen_02}). Major spectral distortions in this range are
excluded by COBE-FIRAS and the parameters of both the
Bose-Einstein and Compton distortions are largely constrained:
$|\mu| < 9 \times 10^{-5}$ and  $|y| < 15 \times 10^{-6}$ (both at
95\% CL, see \cite[]{Fixen_96}).

Although very accurate the FIRAS result covers only part of the
CMB frequency spectrum. In particular it does not cover the low
frequency region where TRIS measurements were made and where
distortions can be expected. Table \ref{tab1} collects the
measurements of the CMB temperature made at frequencies below few
GHz starting from the 80's. At these low frequencies the error
bars are much larger than at FIRAS frequencies, essentially
because of the large uncertainties associated to the absolute
calibration procedures and to the presence of important
foregrounds.

To improve the low frequency situation it has been proposed in the
past to carry on measurements of the CMB temperature from space:
LOBO (\cite[]{Sironi_95}; \cite[]{Sironi_97}) for which TRIS is a
pathfinder and DIMES \cite[]{Kogut_96}. DIMES has been then
transformed in a balloon program (ARCADE), whose results have been
published recently (see \cite[]{Kogut_04}; \cite[]{Fixen_04};
\cite[]{Singal_06}).

The black-body spectrum of the CMB is the result of the complete
thermalization of the Universe in the pre-recombination era. At
that time the strong interactions among the different components
of the cosmic plasma were efficient enough to re-establish
immediately the thermodynamic equilibrium after an energy
injection. Nevertheless small deviations from the black-body
distribution can survive if the energy injection occurred at a
red-shift $z\lesssim 10^6$. In the pre-recombination era
($z\gtrsim 10^4 - 10^3$) a kinetic equilibrium is re-established
through the Compton scattering, Double Compton scattering and the
Bremsstrahlung mechanism. The resulting spectrum assumes a typical
shape described by the Bose-Einstein or by the Compton
distribution. For a review of these mechanisms see
\cite[]{Zeldovich_69}, \cite[]{Sunyaev_70}, \cite[]{Zeldovich_72},
\cite[]{Illarionov_75a},
\\ \cite[]{Illarionov_75b}, \cite[]{Sunyaev_80}, while detailed
calculations have been performed by \cite{Burigana_91a},
\cite{Burigana_91b}, \cite{Burigana_95} and \cite{Daly_91}. These
types of distortions are described by the chemical potential $\mu$
and the Comptonization parameter $y$, whose possible values have
been already limited by COBE-FIRAS.

Energy injections after the recombination era are responsible of
spectral distortions without any possibility of re-thermalization.
Among these processes there is the photon injection through
free-free mechanism in a re-ionized universe (see
\cite[]{Bartlett_91}). This effect depends on the square of the
wavelength and therefore is dominant at very low frequencies and
is not constrained by the COBE-FIRAS measurements. Other possible
distortions are related to the formation of primordial molecules
and to the formation of structures at large scale. Both these
mechanisms are difficult to be modeled but, at least in the case
of the formation of primordial molecules, a not negligible effect
could be expected at decimetric wavelengths (see
\cite[]{Varsha_77} and \cite[]{Dubrovich_95}).

The galactic emission ($T_{gal}(\nu,\alpha,\delta)$) is
anisotropic and has a power law frequency spectrum: $T_{gal}(\nu)
= T_{gal}(\nu_0) (\nu/\nu_0)^{\beta}$. Its importance and
properties are analyzed in the companion paper \cite[]{TRIS-III}
(hereafter Paper III) from which we take the spectral index
$\beta$ we will use in the following.

The contribution of the unresolved extragalactic radio sources
(UERS), $T_{uers}$, has also a power law frequency spectrum, but
for a low resolution experiment it can be considered isotropic. In
the following we will take it from the model developed by
\cite{Gervasi_08} on the basis of the source number counts
measurements available in literature.

After a brief summary of TRIS experiment (Section \ref{TRIS}), in
Section \ref{Calculation} we discuss the techniques used to
separate the components of $T_{sky}$ and the results obtained for
$T_{cmb}$. The implications of these results in terms of spectral
distortions are presented in Section \ref{Discussion}.

\section{TRIS}\label{TRIS}

\begin{figure}
\begin{center}
\epsscale{0.65}\plotone{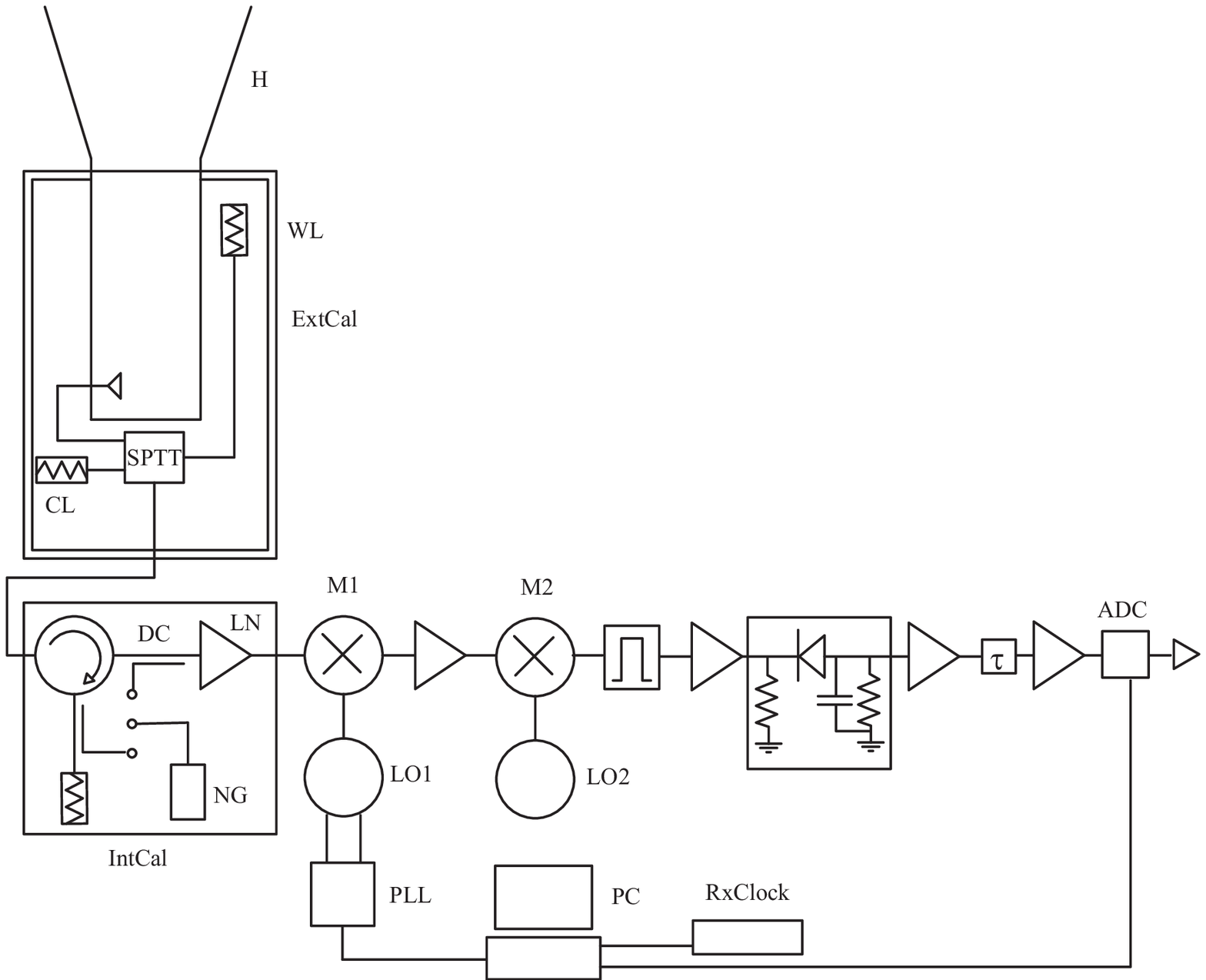} \caption{Schematic of TRIS
antennas and receivers. The set-up of the absolute measurements
using the cryogenic front-end is shown. The internal calibrator
block (IntCal) is also shown. This schematic is the same at the
three frequencies (0.60, 0.82 and 2.5 GHz) of the TRIS experiment.
\label{F1}}
\end{center}
\end{figure}

TRIS experiment, fully described in the companion paper
\cite[]{TRIS-I} (hereafter Paper I), is a set of three
absolute total power radiometers. They were operated (1996-2000)
at Campo Imperatore (Italy), a site at 2000 m a.s.l. where a
reasonable compromise of radio-quietness, winter accessibility and
heavy logistic was achieved. After a few tests to minimize the
effects of the radio-frequency interferences, the radiometers were
tuned at: $\nu = 0.60050$ GHz, $\nu = 0.81785$ GHz, $\nu =
2.42775$ GHz. All the radiometers had the same block diagram (see
Figure \ref{F1}). The antennas, rectangular horns
corrugated on the E plane, were geometrically scaled and had the
same beam size ($18^{\circ}(E)\times23^{\circ}(H)$ HPBW) with
side-lobes lower than -40 db. Two modes of observations were used:
drift scan mode and absolute measurement mode. During the drift
scans the receivers were connected to the horn by a room
temperature waveguide-to-coaxial transition. During absolute
measurements, the radiometers' front-ends were cooled at liquid
helium temperature, as shown in Figure \ref{F1}.

\subsection{Drift Scans}\label{Drift Scans}
Most of the experiment lifetime was used to carry on drift scans
of the sky with the antennas aimed at the zenith
($\delta=+42^\circ$). To avoid sun contamination and keep the gain
variation under control, the sky profiles were built using only
data collected at nighttime, when the temperature stability of the
receivers was better than $\pm 0.1^\circ$C. Receiver gain and
change of the antenna impedance matching were checked at regular
time intervals using the Internal Calibrator (IntCal block in
Figure \ref{F1}), based on a noise source with a
stability better than $\sim 0.1\%$ over months. Data collected in
rainy and cloudy conditions were rejected from the analysis (for
details see Sections 4.2 and 4.4 of the Paper I).

\begin{figure}
\begin{center}
\epsscale{0.5}\plotone{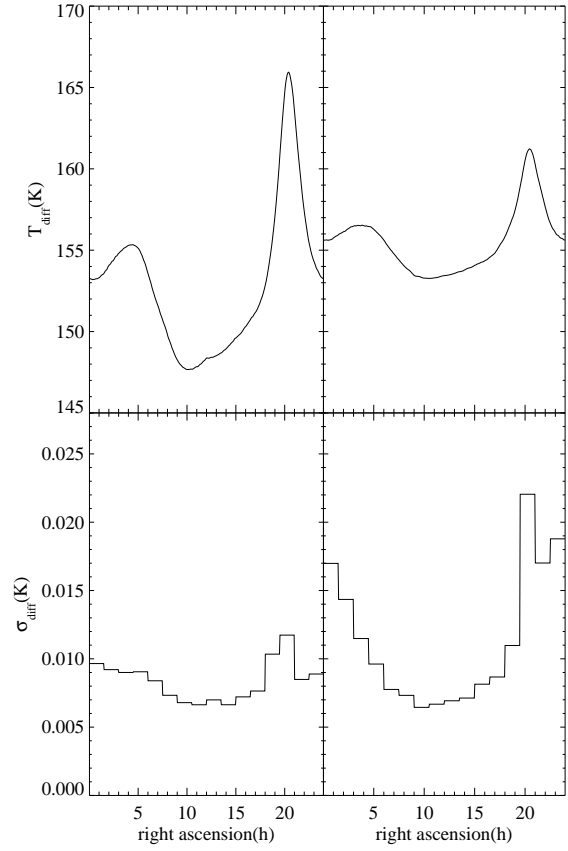} \caption{Drift scan
profiles at 0.60 and 0.82 GHz, with arbitrary zero level. The reported uncertainties represent the statistical
error bar, which is the result of the collection of a large number
of single drift scan measurements. \label{F2}}
\end{center}
\end{figure}

Once properly combined the TRIS drift scans gives complete
profiles of $T_{sky}$ versus the right ascension $\alpha$, along
the circle at declination $\delta=+42^\circ$ at 0.60 and 0.82 GHz.
At 2.5 GHz the level of radio frequency interferencies hampered
the construction of a complete profile. The error budget on the
variations of the temperature along the profiles at 0.60 and 0.82
GHz are dominated by the uncertainties on the corrections for
drifts and offsets and by the uncertainty in the determination of
the temperature scale. In particular, at $8^h \lesssim \alpha
\lesssim 16^h$, the galactic halo region, which has been observed
redundantly, an uncertainty upon temperature variations in the
range from 5 to 10 mK was achieved, both at 0.60 GHz and 0.82 GHz.
At $\alpha \sim 20^h$ (see Figure \ref{F2}), the galactic
disk region, the larger uncertainty we obtained is due to the
smaller number of drift scans. In drift scan mode of observation
(shown in Figure \ref{F2}) the zero level of the scale of
temperature is arbitrary.

\subsection{Absolute Measurements}\label{Absolute Calibrations}
\begin{figure}
\begin{center}
\epsscale{0.65}\plotone{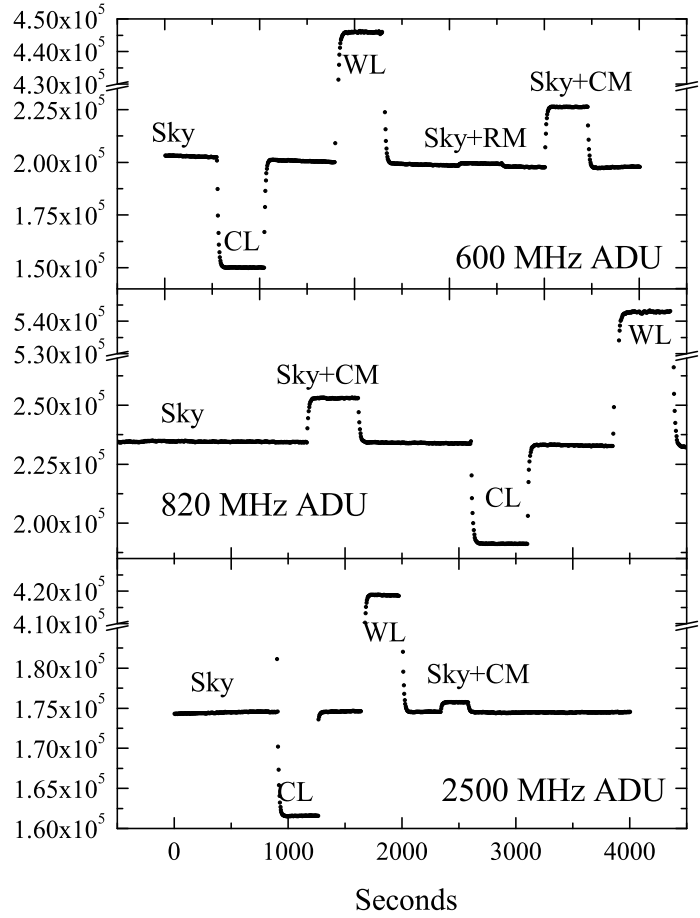} \caption{A typical run
of absolute measurements at the three frequencies of TRIS. We
measured the sky (Sky), the Cold Load (CL), the Warm Load (WL),
the internal calibration mark (Sky+CM) and the impedance matching
(Sky+RM). Signals are still in arbitrary digital units (ADU).
\label{F3}}
\end{center}
\end{figure}

During the absolute measurement campaigns the radiometers were
equipped with the cryogenic front-end. The horns were attached to
stainless steel waveguides and brass waveguide-to-coaxial
transition which were immersed in liquid helium to reduce their
thermal emissions. Each cryostat housed also a Single Pole Triple
Through (SPTT in Figure \ref{F1}) passive resonant
switch. The SPTT connects, for absolute calibration, the receiver
alternately with the antenna or two reference dummy loads: Cold
Load (CL) at $\sim$ 4 K and Warm Load (WL) at $\sim$ 270 K.
Typical calibration runs, involving cold and warm loads and the
internal calibration mark (CM) are shown in Figure \ref{F3}. The transfer function of all the components between the
horn mouth and the switch output (waveguides, transitions, cables,
etc.) have been measured in laboratory. The only parts we were
unable to measure are the flares of the horns: their attenuation
has been calculated. Accurate measurements of the temperature
profiles of these components, both inside and outside the
cryostat, during the observations, combined with the attenuation,
allow to evaluate and subtract their contribution to the receiver
output. At 0.82 GHz a failure of the dewar housing the cryogenic
front-end during absolute measurements forced us to use an
alternative set-up with only the SPTT switch cooled in liquid
Helium. This is the reason why the 0.82 GHz values of the Sky
Temperature have an higher systematic uncertainty (see for details
the Sections 4.2 and 4.3 of the Paper I).

Comparing the signals measured looking at the sky with the signals
produced by the warm and cold loads we get absolute values of the
antenna temperature at various points along the circle at
$\delta=+42^\circ$. Subtracting the environmental contribution
(ground, atmosphere, RF interferences) we get absolute values of
the sky temperature at the same points. Details of this procedure
are given in the Paper I (Section 4.3.1), while results are
presented in Table \ref{tab2}. The ground contribution has been
calculated convolving the antenna beam with a black-body at 290 K
having the profile of the horizon at Campo Imperatore. The
atmospheric transmission and emission have been evaluated using
the model described in \cite[]{Ajello1995}. The effect of RF
interferences have been evaluated by using metal shields around
the horns. At 0.60 and 0.82 GHz they were well inside the system
noise. At 2.5 GHz they were so frequent and strong that
observations were possible only rarely (for details see
Paper I).

The Sky Temperature values ($T_{Sky}$) measured during the
absolute observations were used to set the zero level of the
drift scans. As a matter of fact, the drift scans, once reduced to
the top of the atmosphere, had the same dynamic range of the
absolute measurements data. In this way we got the absolute Sky
Temperature also for those regions of the sky not observed during
the absolute measurements. Table \ref{tab_Tsky} reports the final
accuracy of the absolute values of $T_{Sky}$. The systematic
uncertainty on the zero level quoted in Table \ref{tab_Tsky}
($\Delta T_{zero}$) applies also to the drift scan profiles once
corrected for the zero level. In Table \ref{tab_Tsky} the number
of independent sky positions analyzed during the absolute
measurements is also shown.

\section{Separation of the components of $T_{sky}$}\label{Calculation}

At TRIS frequencies the brightness temperature of the sky can be
written as:

\begin{equation}
T_{sky}(\nu,\alpha) = T_{cmb}(\nu) + T_{gal}(\nu,\alpha) +
T_{uers}(\nu) \label{Tsky_0}
\end{equation}

The contribution ($T_{uers}$) of unresolved extragalactic radio
sources (UERS) at the TRIS frequencies can be obtained from
\cite[]{Gervasi_08} and immediately subtracted from $T_{sky}$. The
values we used are summarized in Table \ref{tab_uers}.

To disentangle $T_{cmb}$ and $T_{gal}$ at 0.60 and 0.82 GHz we
take advantage of the accurate TRIS profiles of $T_{sky}$ versus
$\alpha$ (see Section \ref{600-820}). At 2.5 GHz, because no drift
scan is available, we followed a different approach (see Section
\ref{2.5}).

\subsection{Analysis of the measurements at 0.60 and 0.82 GHz}\label{600-820}

\subsubsection{The position difference technique (PDT)}\label{PDT}

The separation of the components of $T_{sky}$ is possible because
it is made of an isotropic component ($T_{cmb}$ and $T_{uers}$)
plus an anisotropic component ($T_{gal}$). The profiles of
$T_{sky}$ measured by TRIS at 0.60 and 0.82 GHz give the variation
of $T_{sky}$ with the right ascension $\alpha$ along a circle at
constant declination ($\delta=+42^\circ$). We can take a pair of
positions in the sky ($\alpha_1$, $\alpha_2$) at the two
frequencies $\nu_1 =$ 0.60 GHz and $\nu_2 =$ 0.82 GHz, and write a
set of linear equations:

\begin{eqnarray}
T_{sky}(\nu_1,\alpha_1) - T_{uers}(\nu_1) = T_{cmb}(\nu_1) +
T_{gal}(\nu_1,\alpha_1) \nonumber \\
T_{sky}(\nu_1,\alpha_2) - T_{uers}(\nu_1) = T_{cmb}(\nu_1) +
T_{gal}(\nu_1,\alpha_2) \nonumber \\
T_{sky}(\nu_2,\alpha_1) - T_{uers}(\nu_2) = T_{cmb}(\nu_2) +
T_{gal}(\nu_2,\alpha_1) \nonumber \\
T_{sky}(\nu_2,\alpha_2) - T_{uers}(\nu_2) = T_{cmb}(\nu_2) +
T_{gal}(\nu_2,\alpha_2) \label{Tsky_1}
\end{eqnarray}

\noindent The galactic signal depends both on $\alpha$ and $\nu$,
and we can write:

\begin{equation}
T_{gal}(\nu,\alpha) = T_{gal}(\nu_0,\alpha) \ \Bigl(
\frac{\nu}{\nu_0}\Bigr)^{\beta(\alpha,\nu,\nu_0)} \label{gal}
\end{equation}

\noindent In spite of this simple analytical form, also the
spectral index $\beta(\alpha,\nu,\nu_0)$ depends on $\alpha$ and
$\nu$. Putting $\nu_1 = \nu_0$ and taking the difference between
pairs of Equations \ref{Tsky_1} we get:

\begin{eqnarray}
T_{sky}(\nu_1,\alpha_1) - T_{sky}(\nu_1,\alpha_2) =
T_{gal}(\nu_1,\alpha_1) - T_{gal}(\nu_1,\alpha_2) \nonumber \\
T_{sky}(\nu_2,\alpha_1) - T_{sky}(\nu_2,\alpha_2) =
T_{gal}(\nu_1,\alpha_1) \ m(\alpha_1) - T_{gal}(\nu_1,\alpha_2) \
m(\alpha_2) \label{Tsky_2}
\end{eqnarray}

\noindent with \ $m(\alpha) = ( \nu_2/\nu_1 )^{\beta(\alpha)}$. We
can use these equations to separate the microwave sky components
if we can find two positions $\alpha_1$ and $\alpha_2$ such that \
$\beta(\alpha_1) \neq \beta(\alpha_2)$ \ ($m(\alpha_1) \neq
m(\alpha_2)$), \ $T_{sky}(\nu_1,\alpha_1) \neq
T_{sky}(\nu_1,\alpha_2)$ \ and \ $T_{sky}(\nu_2,\alpha_1) \neq
T_{sky}(\nu_2,\alpha_2)$. \ When these conditions, necessary to
break the degeneracy, are satisfied, from Equations \ref{Tsky_2}
follows:

\begin{eqnarray}
[T_{sky}(\nu_2,\alpha_1) - T_{sky}(\nu_2,\alpha_2)] -
[T_{sky}(\nu_1,\alpha_1) - T_{sky}(\nu_1,\alpha_2)] \
m(\alpha_1) = \nonumber \\
= T_{gal}(\nu_1,\alpha_2) \ [m(\alpha_1) - m(\alpha_2)]
\label{Tsky_3}
\end{eqnarray}

\noindent an equation we can use to extract $T_{gal}$, if
$m(\alpha_1)$ and $m(\alpha_2)$ are known. If $m$ is unknown we
can look for different pair of points close to $\alpha_1$ and
$\alpha_2$ respectively, write a system of equations and extract
$T_{gal}(\alpha)$ and $m(\alpha)$. Finally, going back to
Equations \ref{Tsky_1} we can get a number of values of
$T_{cmb}(\nu)$ in the sky regions around $\alpha_1$ and
$\alpha_2$.

\subsubsection{TT-plot technique}\label{TTplot}

For each pair of sky positions $\alpha_i$ and $\alpha_j$, around
$\alpha_1$ and $\alpha_2$ respectively, we can write:

\begin{equation}
m(\alpha_{i,j}) = \frac{T_{sky}(\nu_2,\alpha_i) -
T_{sky}(\nu_2,\alpha_j)}{T_{sky}(\nu_1,\alpha_i) -
T_{sky}(\nu_1,\alpha_j)} \label{eq_TTplot}
\end{equation}

\noindent In principle one can repeat this procedure for all the
pair of points along the drift scan profiles at 0.60 and 0.82 GHz
obtained by TRIS, look for regions where the conditions to breack
the degeneracy are satisfied and get $m$.

This can be done very efficiently using a well known graphical
method, the TT-plot technique, which allows to look for
correlation between the sky temperatures at two frequencies
$\nu_1$ and $\nu_2$. Introduced by \cite{turtle62} and widely used
also recently e.g. in \cite{davies1996} and \cite{platania1998} to
analyze maps of the microwave sky, it allows to find regions where
the values of $m$ (and $\beta$) are well defined. These
"homogeneous" regions can be selected by plotting
$T_{sky}(\nu_2,\alpha)$ versus $T_{sky}(\nu_1,\alpha)$: all the
positions $\alpha$ inside these regions distribute themselves
along a straight line whose slope is $m$. This means that in these
positions the spectral index can be considered uniform:
$\beta(\alpha_i) \simeq \beta(\alpha_j)$.

As evident from Equation \ref{eq_TTplot} the values of $m(\alpha)$
(and $\beta(\alpha)$) obtained depend only on the temperature
variations and are not affected by uncertainties on the zero level
of the two scales of temperature ($\Delta T_{zero}$). In the
Paper III \cite[]{TRIS-III}, Section 3, the TT-plot method
has been applied to the TRIS profiles at 0.60 and 0.82 GHz (shown
in Figure \ref{F2}). It shows that, along the circle at
$\delta = +42^\circ$, there are two regions where the spectral
indices are well defined and different: here the conditions to
solve Equation \ref{Tsky_3} apply. These two regions are: \ $9^h
\lesssim \alpha_1 \lesssim 11^h$, \ the halo region at the highest
galactic latitude; and \ $20^h \lesssim \alpha_2 \lesssim 21^h$, \
the galactic disk, close to the Cygnus region.

\subsubsection{Monte Carlo analysis}\label{MC}

Having the above information we can now solve the equation and get
$T_{cmb}$ and $T_{gal}$ at various $\alpha$. We did it by a Monte
Carlo simulation. In this way we take directly into account the
effects of the uncertainties, both statistic and systematic, of
the measured quantities ($T_{sky}(\nu,\alpha)$) on the evaluation
of the unknown quantities. We also assumed as prior the
information coming from the TT-plot method used in Paper
III: i.e. central value and width of the distribution of the
spectral index $\beta(\alpha)$. Following this approach we
obtained the distribution of the values of $T_{cmb}(\nu_1)$ and
$T_{cmb}(\nu_2)$, which include the effect of the distribution of
all the parameters involved in the Equations \ref{Tsky_1}. The
error bars of $T_{cmb}$ have been evaluated by using the bootstrap
method. The systematic uncertainty of the zero level of $T_{sky}$
($\Delta T_{zero}$ in Table \ref{tab_Tsky}), not included in the
MC analysis, has been quoted separately.

\begin{figure}
\begin{center}
\epsscale{0.5}\plotone{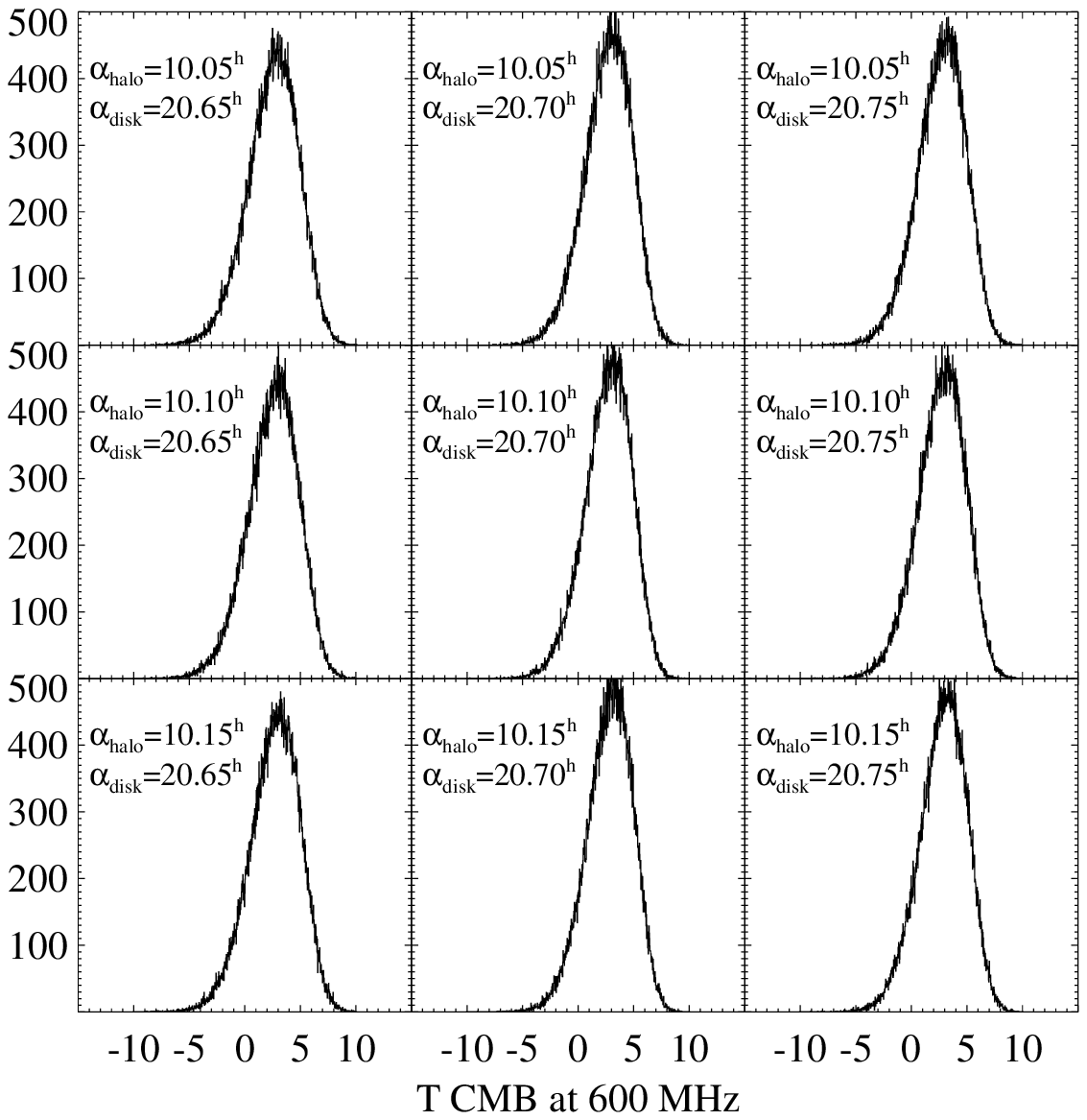}
\epsscale{0.5} \plotone{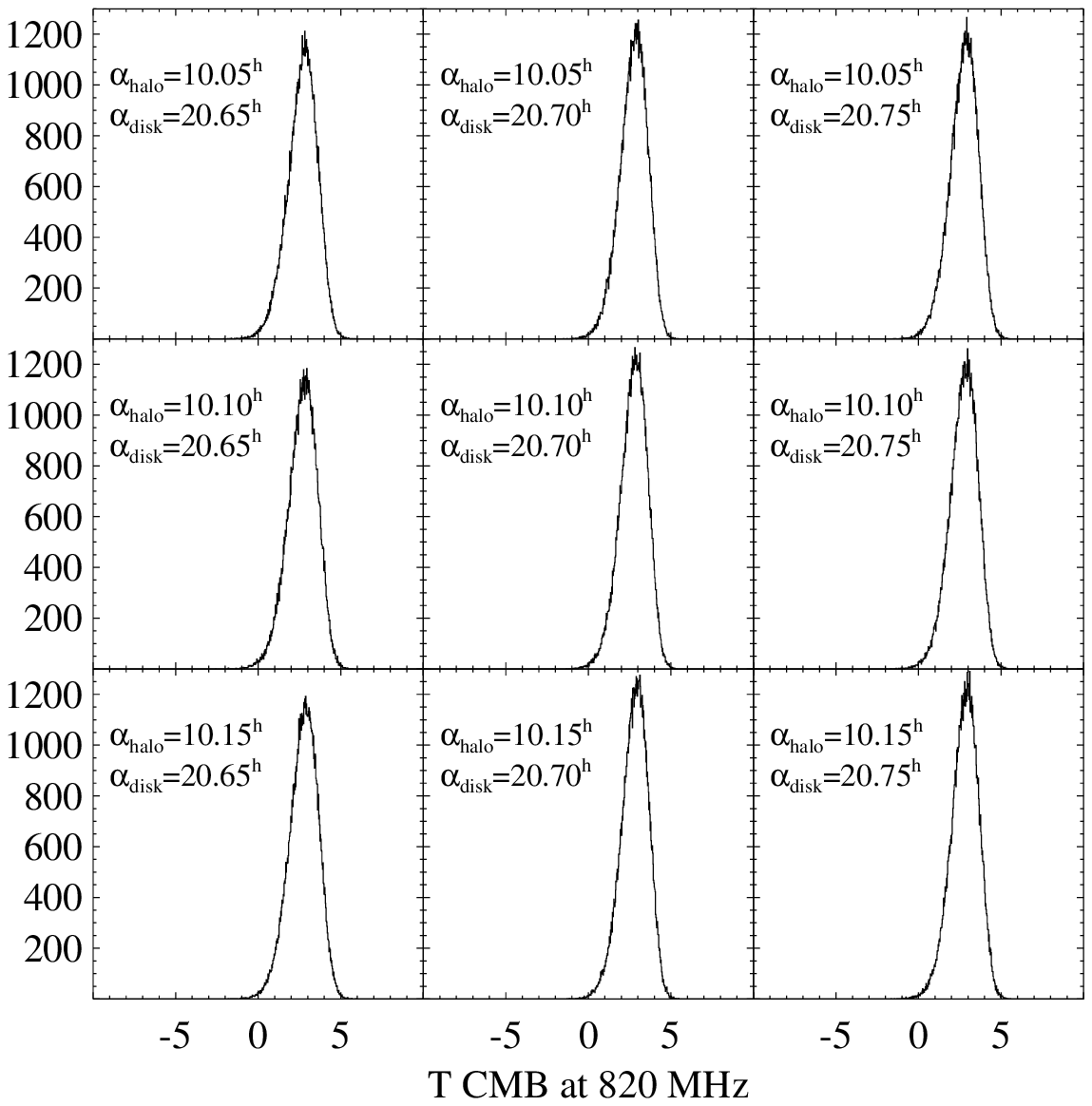}
\caption{Brightness temperature of CMB evaluated with the position
difference technique at 0.60 and 0.82 GHz. We got values from 288
pairs of position in the sky (32 positions in the halo $\times$ 9
positions in the disk). Here we show the Monte Carlo realizations in 9
samples of sky position pairs. \label{F4}}
\end{center}
\end{figure}

The values of $T_{cmb}$ we obtained are distributed around a well
defined central value, as shown in Figure \ref{F4} and
summarized in Table \ref{tab_cmb-TTp}. We have also obtained the
galactic emission in the halo and in the disk regions (see Table
\ref{tab_gal}). As shown by the system of Equations \ref{Tsky_2},
the absolute temperature of the galaxy, evaluated using this
method, is not affected by the systematic uncertainty ($\Delta
T_{zero}$) on the zero level of $T_{sky}$.

\subsubsection{Graphic method}\label{Intercetta}

In principle graphic methods can be used also to extract $T_{cmb}$
from a TT-plot. Through each point ($T_{sky}(\nu_1,\alpha)$;
$T_{sky}(\nu_2,\alpha)$), at a certain value of $\alpha$, of the
TT-plot we can draw a straight line of slope $m(\alpha) = \bigl(
\nu_2/\nu_1 \bigr)^{\beta(\alpha)}$ whose intercept $T_0$ is:

\begin{equation}
T_0 = T_{sky}(\nu_1,\alpha) - T_{sky}(\nu_2,\alpha) \
\frac{1}{m(\alpha)} \label{interc}
\end{equation}

\noindent After a simple algebra, and assuming $T_{cmb}(\nu_1)
\equiv T_{cmb}(\nu_2) = \overline{T_{cmb}}$, we get:

\begin{equation}
\overline{T_{cmb}} = \frac{m}{m-1} \Bigl[ T_{sky}(\nu_1) -
T_{sky}(\nu_2) \ \frac{1}{m} - T_{uers}(\nu_1)
\Bigl(1-\frac{q}{m}\Bigr) \Bigr] \label{interc2}
\end{equation}

\noindent with $q = T_{uers}(\nu_2)/T_{uers}(\nu_1)$. We have used
this method in the same halo and disk regions used before. In this
way we got values of $\overline{T_{cmb}}$ in agreement with the
results obtained by the PDT method. However: (i) we can not
separate the CMB temperature at the two frequencies and (ii) there
is an unfavorable combination of systematics on the zero level
assessment at the two frequencies:

\begin{equation}
\Delta T^{CMB}_{zero} = |\frac{m}{m-1}| \ \Delta T^{SKY}_{zero}
(\nu_1) + |\frac{1}{m-1}| \ \Delta T^{SKY}_{zero} (\nu_2)
\end{equation}

\noindent which, applied to TRIS data at $\nu_1=0.60$ and
$\nu_2=0.82$ GHz, gives \ $m \simeq 0.38-0.42$ and $\Delta
T^{CMB}_{zero} =^{+0.76}_{-0.54}$ K. In order to avoid these
limitations we need to consider differences between pairs of
positions in the sky. But in this case the set of equations we get
is equivalent to the PDT method described above.

\subsection{Analysis of the measurements at 2.5 GHz}\label{2.5}

At 2.5 GHz no profile of $T_{sky}$ vs $\alpha$ was available.
Therefore we were forced to adopt a more straightforward but less
accurate method: evaluating $T_{gal}$ at $\nu=2.5$ GHz from
independent measurements, by a model, and subtracting it from \
$T_{sky} - T_{uers}$ \ to get $T_{cmb}$. We got $T_{gal}$
extrapolating data from the map of the diffuse radiation at 1.42
GHz prepared by \cite{Reich_86}, which is the one in literature
closest to $\nu=2.5$ GHz, covering the sky region scanned by TRIS.
This map has been convolved with the beam of the TRIS antennas in
order to get the synthetic drift scan at $\delta =+42^\circ$. Then
we extrapolated this signal at 2.5 GHz using the spectral index
$\beta$ got from TRIS data at 0.60 and 0.82 GHz (see Paper III).
Finally we subtracted
$T_{gal}$ and $T_{uers}$ from $T_{sky}$ and got $T_{cmb}$. The
results are reported in Table \ref{tab_cmb-25}. Here the
statistical uncertainty on the experimental points ($\Delta
T_{stat}=103$ mK) is larger than at lower frequencies because we
have few independent measurements.

\subsection{Results}\label{Results}

The values of $T_{cmb}$ obtained above, using different methods,
are consistent within the error bars (see Section \ref{600-820}).
This means that the errors (statistics and systematics), proper of
the measurements and arising from the analysis, are well
represented by the quoted uncertainty (see Table
\ref{tab_cmb-TTp}).

\begin{figure}
\epsscale{1.0} \plotone{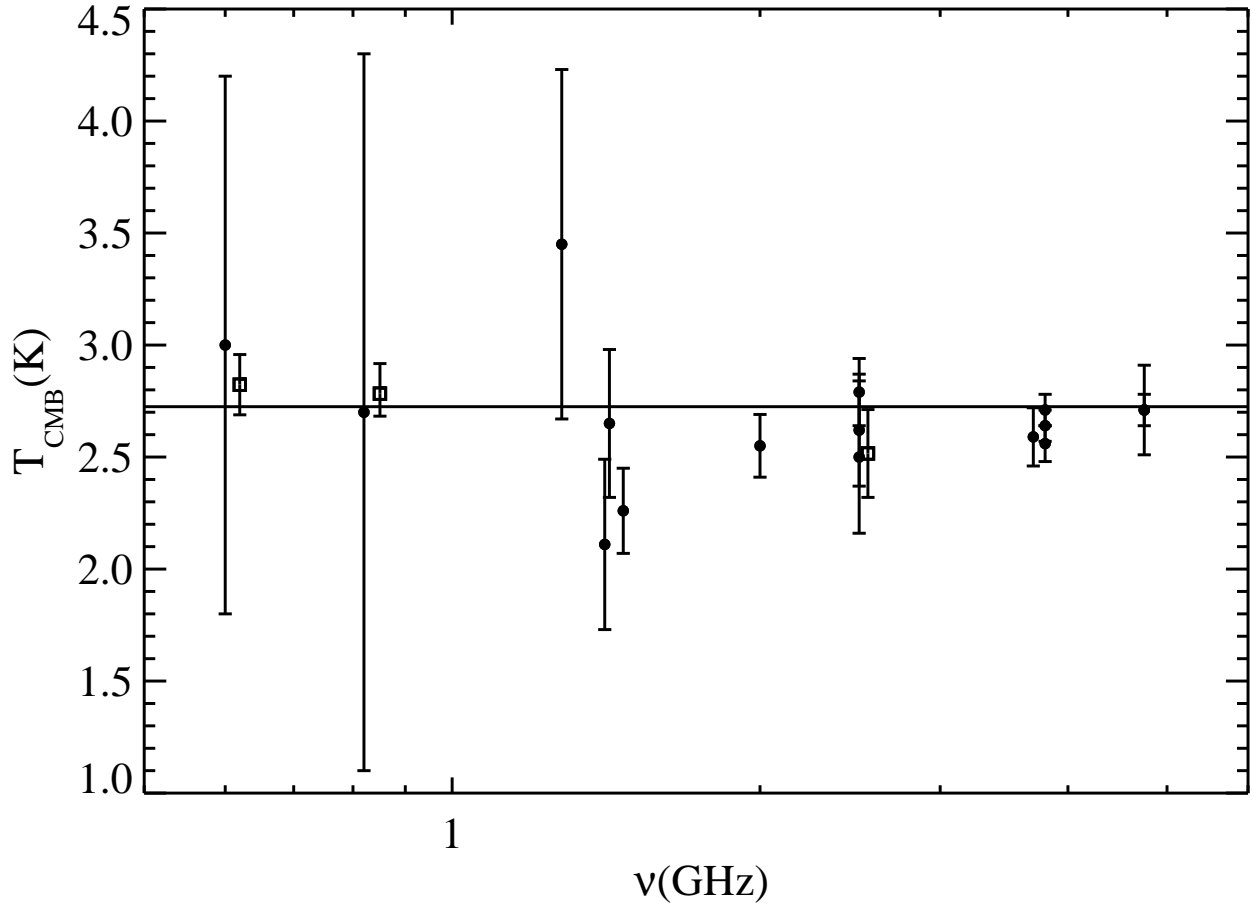} \caption{The CMB thermodynamic
temperature measured at low frequencies (see Table \ref{tab1}).
For easier comparison with previous measurements (solid circles),
TRIS data points (open squares) have been slightly shifted in
frequency. The horizontal solid line is the CMB temperature
obtained by FIRAS at higher frequencies. \label{F5}}
\end{figure}

The brightness temperatures of the CMB ($T_{cmb}(\nu)$) are
converted into thermodynamic black-body temperature
($T_{cmb}^{th}(\nu)$):

\begin{equation}
T_{cmb}^{th}(\nu) = \frac{h \nu}{k \ln(x_b + 1)} \label{Tbb}
\end{equation}

\noindent where \ $x_b = h \nu / k T_{cmb}$. Results are
summarized in Table \ref{tab_cmb-BB} and in Figure \ref{F5}
together with the results from previous measurements (see Table
\ref{tab1}).

\section{Discussion}\label{Discussion}

\subsection{Comparison with the previous measurements}

The results obtained by the TRIS experiment represent a
significant improvement in the CMB measurements at frequencies
lower than 1 GHz. 

Above 1 GHz in literature there are more accurate and recent
results. These measurements give temperatures of the CMB which,
within the error bars, are in agreement with the temperature
measured by FIRAS at much higher frequencies. The only exceptions
are the results obtained at 1.4 GHz by \cite{Levin_88} and
\cite{Bensadoun_93}. It is hard to fit these results only with the
spectral distortions allowed by COBE-FIRAS (see
\cite[]{Fixen_96}).

Among the more recent experiments, the most promising program is
ARCADE (see \cite[]{Kogut_03}). At the frequencies $\nu = 10$ and
30 GHz ARCADE has obtained results with error bars of 10 and 32
mK respectively (see \cite[]{Fixen_04}), which at 8.0 and 8.3 GHz
go up to 120 and 160 mK respectively (see \cite[]{Singal_06}). The
lowest frequency scheduled by ARCADE is 3.3 GHz. Unfortunately
above $2-3$ GHz an accuracy of few mK is necessary to detect or
furtherly constrain the limits on distortions produced by
Comptonization, Bose-Einstein (\emph{BE}), or free-free
(\emph{FF}) processes, set by FIRAS. The search for \emph{BE} and
\emph{FF} distortions is better done below $2-3$ GHz. Here in fact
larger distortions can be expected.

We can summarize our results in the following way:

\noindent \emph{(I)} At $\nu = 0.60$ GHz we reduced the error bar
by a factor $\sim 9$. Before TRIS the error bar was equally
distributed between uncertainties on the temperature of the sky
and uncertainties on the level of the galactic and extragalactic
foregrounds (see \cite[]{Sironi_90}). The better accuracy of TRIS
is due to improvement of the experimental set-up and better
procedure of foregrounds subtraction.

\noindent \emph{(II)} At $\nu = 0.82$ GHz the error bar reduction
resulting from TRIS is a factor $\sim 7$. Due to the failure of
the calibrator TRIS results are are still dominated by the
systematic uncertainty on the zero level, while the error bar due
to the foregrounds separation is smaller.

\noindent \emph{(III)} The TRIS measurement at $\nu = 2.5$ GHz
does not improve but is fully compatible with the previous results
(see Table \ref{tab1}). Combined with all the observations in
literature we get the value $T_{cmb}^{th}(2.5 \ GHz) = 2.680 \pm
0.110$ K.

\subsection{The spectral distortions}

We can now combine TRIS results with the data in literature and
look for better limits on spectral distortions. The Compton
distortions are important only in the Wien region of the CMB
spectrum. Therefore in the following discussion we will not
consider them.

The Bose-Einstein (\emph{BE}) distortion is, conversely, important
at frequencies lower than 1 GHz. Various analyses suggest it can
produce a dip in the brightness temperature at sub-GHz frequencies
(see \cite[]{Burigana_95}; \cite[]{Burigana_03}). The minimum is
expected at a frequency:

\begin{equation}
\nu_{min}^{BE} \propto [\Omega_b h^2]^{2/3} \label{BEmin}
\end{equation}

\noindent where $\Omega_b$ is the barion density and $h$ is the
Hubble constant parameter. Assuming the WMAP results for $\Omega_b
h^2$ (see \cite[]{Spergel_07}) we expect $\nu_{min}^{BE} \sim
0.3-0.4$ GHz. Detection of $\nu_{min}^{BE}$ should give an
independent estimate of the baryon density $\Omega_b$. The dip
amplitude is related to the chemical potential $\mu$, by \ $\Delta
T_{BE} \propto \mu \ [\Omega_b h^2]^{-2/3}$. The limit set by
COBE-FIRAS on the chemical potential ($|\mu| < 9 \times 10^{-5}$)
implies a spectral distortion $\Delta T_{BE} \lesssim 17$ mK (95\%
CL), so the current measurements at low frequency, including TRIS,
are not yet accurate enough to detect $\nu_{min}^{BE}$ and
constrain furtherly $\Omega_b$.

The free-free emission (\emph{FF}) is expected to produce a
temperature deviation with a quadratic dependence on the
wavelength:

\begin{equation}
\Delta T_{ff} = T_0 \ \frac{Y_{ff}}{x^2} \label{dTFF}
\end{equation}

\noindent where $Y_{ff}$ is the optical depth to free-free
emission and $x$ is the  dimensionless frequency (see
\cite[]{Bartlett_91}). This kind of distortion is expected in case
of re-ionization of the intergalactic medium. The amplitude of the
cosmological signal depends on the integrated column density of
ionized gas produced at the redshift of formation of the first
collapsed objects \cite[]{Kogut_03} and on the thermal history of
the IGM through the electrons' temperature $T_e(z)$. The current
upper limit set by the previous measurements at low frequency is
$|Y_{ff}| < 1.9 \times 10^{-5}$ \cite[]{Bersanelli_94}. A lower
limit can be set by the observed Lyman-$\alpha$ forest: $|Y_{ff}|
\gtrsim 8 \times 10^{-8}$ and $\Delta T_{ff} \gtrsim 2-3$ mK at
$\nu = 0.6$ GHz \cite[]{Haiman_97}. A distortion $\Delta T_{ff}
\sim 30$ mK at $\nu = 0.6$ GHz, corresponding to $|Y_{ff}| \sim
1.3 \times 10^{-6}$ due to a clumpy component from halos has been
suggested by \cite{Oh_99}. Moreover several reionization models
have been studied by \cite{Weller_99}, where values of $ |Y_{ff}|
\sim (0.5 - 4) \times 10^{-8}$ are suggested for different
scenarios.

\begin{figure}
\epsscale{1.0} \plotone{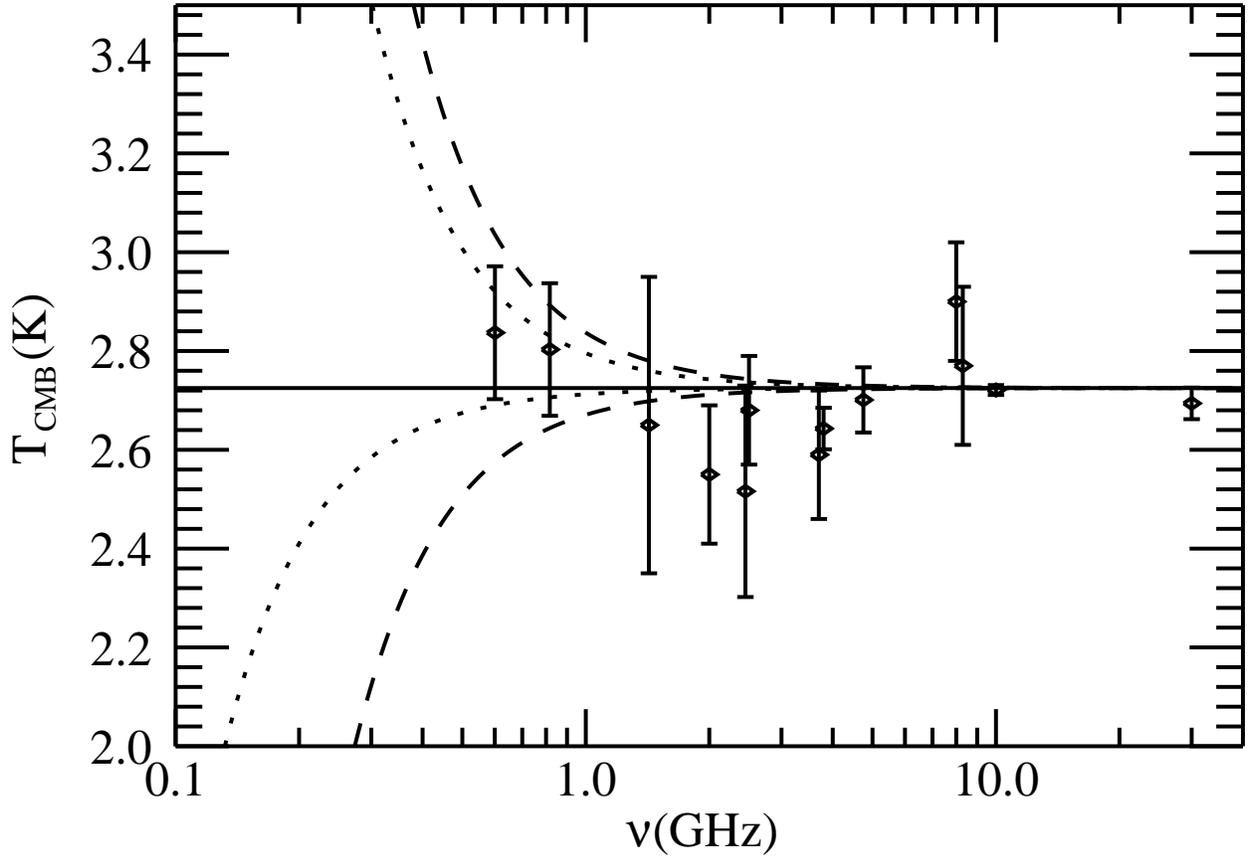} \caption{Distorted spectra
due to \emph{FF} effect and measured CMB temperatures. Short dashed lines correspond to
1$\sigma$ upper limits, while long dashed lines correspond to
2$\sigma$ upper limits (95\% CL). The solid line is the
undistorted CMB temperature. Measurements at the same frequencies
have been combined in this plot. Here we report the error bars
quoted in literature for the various experiments. TRIS data error
bars (1$\sigma$) result from the combination of statistical and
systematic uncertainties. \label{F6}}
\end{figure}

Combining TRIS measurements and the data in literature we get a
data set which has been used to fit the distorted (\emph{FF} +
\emph{BE}) spectra of $T_{cmb}$. It includes all the measurements
reported in the Table \ref{tab1}, plus the COBE-FIRAS data
\cite[]{Fixen_96} and the measurements made by ARCADE
(\cite[]{Fixen_04}; \cite[]{Singal_06}). We have combined the
uncertainties of TRIS data assuming for the statistical one
($\sigma$) a normal distribution and for the systematic one
($\Delta T_{zero}$) a uniform distribution. We set new
limits to \emph{FF} effect: $ -6.3 \times 10^{-6} < Y_{ff} < 12.6
\times 10^{-6}$ at 95\% confidence level. We also improved
marginally the upper limits to \emph{BE} distortions: $|\mu| < 6
\times 10^{-5}$ at 95\% CL. The undistorted black-body temperature
has also been fitted and we have confirmed, as expected, the
results obtained by \cite{Fixen_02}. In Figure \ref{F6} we show
the maximum frequency distortions, due to \emph{FF} effect,
allowed after the present analysis. As clearly shown in Figure
\ref{F6}, the major improvement in the upper limits of $Y_{ff}$
is due to the TRIS measurements, in particular the one at 0.60
GHz. The limits on $Y_{ff}$ cover positive and negative values.
The sign depends on the difference between the electron
temperature $T_e$ in the ionized medium and the radiation
temperature $T_{\gamma}$. Mechanisms of recombination cooling can
in fact lower the electron temperature down to $T_e \gtrsim 0.2
T_{\gamma}$ corresponding to $Y_{ff} \gtrsim -2.3 \times 10^{-5}$
(see \cite[]{Stebbins_86}).

\subsection{Accuracy and perspectives}

The total error budget of the CMB temperature measured by TRIS can
be roughly separated in two blocks: uncertainty in the absolute
measurements ($\Delta T_{zero}$); uncertainty due to the procedure
of foregrounds subtraction ($\sigma$). For TRIS data at 0.60 and
0.82 GHz the statistics is not the dominant souce of uncertainty.
At 2.5 GHz the results are limited also by the statistics.

Systematic uncertainties on the zero level are set by the
capability of controlling and measuring losses and temperatures of
the front-end components. As discussed in Paper I at 0.60
GHz we get an accuracy $\Delta T_{zero} = 66$ mK, a value
achievable also at 0.82 GHz (we could not reach it due to
the failure of the cryogenic system).

The evaluation of the foregrounds and their subtraction from
$T_{sky}$ is the other main source of uncertainty. The UERS
brightness temperature can be evaluated by analyzing the number
counts measurements available in literature (see
\cite[]{Gervasi_08}). Improvements on this value can only come
from better source count measurements. The galactic emission has
been obtained as a byproduct of TRIS observations by the TT plot method. A technique
which works properly if the signals are taken by identical
instruments at different wavelengths. The use of existing maps is
hampered by intercalibration problems, pointing errors, beam shape
accuracy, etc. Minor source of uncertainty are the atmospheric and
ground contribution to the antenna temperature. These terms can be
controlled by a proper design of the experiment.

We estimate that the best accuracy achievable at TRIS frequencies
for $T_{cmb}$, using the TRIS methods, is $\Delta T_{tot} \sim
100$ mK. Better results require to cool down the full front-end,
including the horn, and use a cryogenic calibrator in front of it.
This is the strategy adopted by ARCADE, but this experiment is
limited to frequencies $\nu \gtrsim 3$ GHz (see \cite[]{Fixen_04}
and \cite[]{Singal_06}). Probably multifrequency space experiments
at long wavelengths (see for example proposals like LOBO
(\cite[]{Sironi_95}; \cite[]{Sironi_97}) and DIMES
\cite[]{Kogut_96}), taking advantage of the very stable conditions
available in space, should allow to bring $\Delta T_{tot}$ below
100 mK also at frequencies lower than 1 GHz.

\section{Conclusions}

Starting from the absolute measurements of the sky brightness
temperature, performed by the TRIS experiment, and presented in
Paper I \cite[]{TRIS-I}, we have evaluated the absolute
temperature of the Cosmic Microwave Background at $\nu =$ 0.60,
0.82 and 2.5 GHz.

The thermodynamic temperatures of the CMB we get are:
$T_{cmb}^{th}=(2.837 \pm 0.129 \pm 0.066)K$ at $\nu=0.60$ GHz;
$T_{cmb}^{th}=(2.803 \pm 0.051 \ ^{+0.430} _{-0.300})K$ at
$\nu=0.82$ GHz; $T_{cmb}^{th}=(2.516 \pm 0.139 \pm 0.284)K$ at
$\nu=2.5$ GHz. The first error bar is 1$\sigma$ statistics, while
the second one is the systematic on the zero level assessment.

Thanks to improvements of the absolute calibration system and in
the foregrounds separation technique TRIS succeded in reducing
previous uncertainties by a factor $\sim 9$ at $\nu=0.60$ GHz and
by a factor $\sim 7$ at $\nu=0.82$ GHz. At 2.5 GHz TRIS results
are in agreement with the previous measurements.

These results, used to look for CMB spectral distortions, give an
upper limit to the chemical potential $|\mu| < 6 \times 10^{-5}$
(95\% CL) used to describe the \emph{BE} distortions. We have also
constrained the free-free distortions to: $ -6.3 \times 10^{-6} <
Y_{ff} < 12.6 \times 10^{-6}$ (95\% CL), approaching the values
suggested by observations of the Lyman-$\alpha$ forest
\cite[]{Haiman_97}.

\acknowledgments

{\bf Acknowledgements}: The TRIS activity has been supported by
MIUR (Italian Ministry of University and Research), CNR (Italian
National Council of Research) and the Universities of Milano and
of Milano-Bicocca. The logistic support at Campo Imperatore was
provided by INFN, the Italian Institute of Nuclear Physics, and
its Laboratorio Nazionale del Gran Sasso. The authors acknowledge
also an anonimous referee for his comments which helped us to
improve the quality of the results presented.

\vfill \eject

\begin{deluxetable}{lllllll}
\tablecolumns{7} \tablewidth{0pc} \tablecaption{A summary of low
frequency CMB absolute temperature measurements collected starting
from the 80's. \label{tab1}} \tablehead{ \colhead{$\lambda$ (cm)}
& \colhead{} & \colhead{$\nu$ (GHz)} & \colhead{} &
\colhead{$T_{cmb}$ (K)} & \colhead{} & \colhead{Reference} }
\startdata
50.0  & & 0.60  & & $3.0  \pm 1.2$  & & \cite{Sironi_90} \\
36.6  & & 0.82  & & $2.7  \pm 1.6$  & & \cite{Sironi_91} \\
23.4  & & 1.28  & & $3.45 \pm 0.78$ & & \cite{Raghunathan_00} \\
21.3  & & 1.41  & & $2.11 \pm 0.38$ & & \cite{Levin_88} \\
21.05 & & 1.425 & & $2.65 \ ^{+0.33}_{-0.30}$ & & \cite{Staggs_96} \\
20.4  & & 1.47  & & $2.26 \pm 0.19$ & & \cite{Bensadoun_93} \\
15.0  & & 2.0   & & $2.55 \pm 0.14$ & & \cite{Bersanelli_94} \\
12.0  & & 2.5   & & $2.62 \pm 0.25$ & & \cite{Sironi_84} \\
12.0  & & 2.5   & & $2.79 \pm 0.15$ & & \cite{Sironi_86} \\
12.0  & & 2.5   & & $2.50 \pm 0.34$ & & \cite{Sironi_91} \\
8.1   & & 3.7   & & $2.59 \pm 0.13$ & & \cite{Deamici_88} \\
7.9   & & 3.8   & & $2.56 \pm 0.08$ & & \cite{Deamici_90} \\
7.9   & & 3.8   & & $2.71 \pm 0.07$ & & \cite{Deamici_90} \\
7.9   & & 3.8   & & $2.64 \pm 0.07$ & & \cite{Deamici_91} \\
6.3   & & 4.75  & & $2.71 \pm 0.20$ & & \cite{Mandolesi_84} \\
6.3   & & 4.75  & & $2.70 \pm 0.07$ & & \cite{Mandolesi_86} \\
\enddata
\end{deluxetable}

\begin{deluxetable}{lrrc}
\tablewidth{0pt} \tablecaption{Local contribution to the antenna
temperature in the TRIS experiment. Atmospheric emission was
evaluated by using a model \cite[]{Ajello1995}; ground contribution
was evaluated convolving the ground profile with the antenna
side-lobes; RF interferences were directly measured.}
\tablecolumns{4} \tablehead{\colhead{$\nu_{0}$} & \colhead{0.60 GHz}
& \colhead{0.82 GHz} & \colhead{2.5 GHz}} \startdata
$ T_{atm} $ (K) & $ 1.088 \pm ~0.023 $ & $ 1.221 \pm 0.016 $ & $ 1.570 \pm 0.025 $ \\
$ e^{-\tau_{atm}} $ & $ 0.995 \pm 0.001 $ & $ 0.995 \pm 0.001 $ & $ 0.993 \pm 0.001 $\\
$ T_{ground} $ (K) & $ 0.07^{+0.06}_{-0.03} $ & $ 0.07^{+0.06}_{-0.03} $ & $ 0.07^{+0.06}_{-0.03} $ \\
$ T_{rfi} $ (K) & $ <0.01 $ & $ <0.01 $ & $ 9.82 \pm 0.26 $ \\
\enddata
\label{tab2}
\end{deluxetable}

\begin{deluxetable}{lccc}
\tablewidth{0pt} \tablecaption{Accuracy of the absolute
measurements of $T_{sky}$ performed by TRIS at $\delta = +
42^{\circ}$.} \tablecolumns{4} \tablehead{\colhead{$\nu $ (GHz)} &
\colhead{$\Delta T_{stat}$ (mK)} & \colhead{$\Delta T_{zero}$
(mK)} & \colhead{\# sky positions} } \startdata
0.60 & 18 & 66 & 34 \\
0.82 & 32 & $^{+430}_{-300}$\tablenotemark{\dag} & 12 \\
2.5 & 10 & 284 & 6 \\
\enddata
\tablenotetext{\dag}{ The zero level uncertainty at 0.82 GHz
coming from the absolute calibration is $\Delta T_{zero} = 659$
mK. The quoted value comes after astrophysical assumption on the
galactic signal (see Paper III \cite[]{TRIS-III}, Section
3).} \label{tab_Tsky}
\end{deluxetable}

\begin{deluxetable}{lll}
\tablecolumns{5} \tablewidth{0pc} \tablecaption{Contribution of
the unresolved extragalactic radio sources to the sky brightness
temperature at the frequencies of the TRIS experiment.
\label{tab_uers}} \tablehead{ \colhead{$\nu $ (GHz)} & \colhead{}
& \colhead{$T_{uers}$ (mK)}} \startdata
\ \ \ 0.60 & & \ \ \ \ $934  \pm 24$ \\
\ \ \ 0.82 & & \ \ \ \ $408  \pm  10$ \\
\ \ \ 2.50 & & \ \ \ \ $22   \pm  1$ \\
\enddata
\end{deluxetable}

\begin{deluxetable}{lcccccccccccc}
\tablecolumns{4} \tablewidth{0pc} \tablecaption{Brightness
temperature of CMB at 0.60 and 0.82 GHz evaluated using the PDT
method and the spectral index $\beta$ obtained with the TT-plot
technique as prior in the Monte Carlo simulations. The error bar
due to the systematics (zero level assessment) is quoted
separately from the rest of the uncertainty. \label{tab_cmb-TTp}}
\tablehead{\colhead{$\nu $ (GHz)} & \colhead{} &
\colhead{$T_{cmb}(K)$} & \colhead{} & \colhead{$\Delta
T_{zero}(K)$} & \colhead{} & \colhead{$\Delta T_{MC}(K)$} }
\startdata
0.60 & & 2.823 & & 0.066 & & 0.129 \\
0.82 & & 2.783 & & $^{+0.430}_{-0.300}$ & & 0.051 \\
\enddata
\end{deluxetable}

\begin{deluxetable}{lrrrrrr}
\tablecolumns{10} \tablewidth{0pc} \tablecaption{Brightness
temperature of the galaxy around the minimum and the maximum,
respectively, as measured by TRIS antennas at $\delta =
+42^{\circ}$. In the last column the related spectral index is
reported. \label{tab_gal}} \tablehead{\colhead{R A} & \colhead{} &
\colhead{0.60 GHz} & \colhead{} & \colhead{0.82 GHz} & \colhead{} & \colhead{} \\
\cline{1-1} \cline{3-3} \cline{5-5} \\
\colhead{h} & \colhead{} & \colhead{$T_{gal}^{TRIS}(K)$} & \colhead{} &
\colhead{$T_{gal}^{TRIS}(K)$} & \colhead{} & \colhead{$\beta$} }
\startdata
10.0 (Halo)& & $5.72 \pm 0.07$ & & $2.21 \pm 0.03$ & & $3.09 \pm 0.15$ \\
20.4 (Disk)& & $24.44 \pm 0.07$ & & $10.38 \pm 0.03$ & & $2.76 \pm 0.10$ \\
\enddata
\end{deluxetable}

\begin{deluxetable}{lcccccccccc}
\tablecolumns{9} \tablewidth{0pc} \tablecaption{Brightness
temperature of CMB at 2.5 GHz. The error budget due to the
different contributions is quoted separately. Due to the small
number of independent measurements, the statistical uncertainty of
the experimental points is not negligible. \label{tab_cmb-25}}
\tablehead{\colhead{$\nu $ (GHz)} & \colhead{} &
\colhead{$T_{cmb}(K)$} & \colhead{} & \colhead{$\Delta
T_{zero}(K)$} & \colhead{} & \colhead{$\Delta T_{stat}(K)$} &
\colhead{} & \colhead{$\Delta T_{gal}(K)$} & \colhead{} &
\colhead{$\Delta T_{uers}(K)$}} \startdata
2.50 & & 2.458 & & 0.284 & & 0.103 & & 0.093 & & 0.001 \\
\enddata
\end{deluxetable}

\begin{deluxetable}{lcccccc}
\tablecolumns{7} \tablewidth{0pc} \tablecaption{Thermodynamic
temperature of CMB at 0.60, 0.82 and 2.5 GHz. Systematic and
combined statistic ($1\sigma$) error bar are quoted separately.
\label{tab_cmb-BB}} \tablehead{\colhead{$\nu $ (GHz)} & \colhead{}
& \colhead{$T_{cmb}^{th}(K)$} & \colhead{} & \colhead{$\Delta
T_{zero}(K)$} & \colhead{} & \colhead{$\sigma (K)$}} \startdata
0.60 & & 2.837 & & 0.066 & & 0.129 \\
0.82 & & 2.803 & & $^{+0.430}_{-0.300}$ & & 0.051 \\
2.50 & & 2.516 & & 0.284 & & 0.139 \\
\enddata
\end{deluxetable}

\end{document}